\documentstyle[sprocl,epsf,cite,wrapfig]{article}

\newcommand{\slashed}[1]{\rlap{$#1$}/}
\newcommand{\slashp}{\mbox{$\not \hspace*{-1.10mm} p$}}

\newcommand{\GeV}{\mbox{\rm GeV}}

\newcommand{\lsim}[1]{
\setlength{\unitlength}{12pt}
\begin{picture}(1.4,1.)
\put(.7,-0.3){\makebox(0.0,1.)[t]{$<$}}
\put(.7,-0.3){\makebox(0.0,1.)[b]{$\sim$}}
\end{picture}#1}
\newif\ifpreprint
\preprinttrue

\ifpreprint
\fi

\begin{document}

\title{
The $\gamma^\star\gamma\to\pi^0$ transition in a bound-state
approach
\ifpreprint
   \footnote{The talk given by D. Klabu\v{c}ar at Semi-Exclusive 
Workshop, Jefferson Lab, May 19--22, 1999. 
To appear in the proceedings of the Workshop.}
\fi
      }

\author{DUBRAVKO KLABU\v{C}AR}
 
\address{Physics Department, Faculty of Science,
        University of Zagreb,  \\
Bijeni\v{c}ka c. 32, Zagreb, Croatia}
 
\author{DALIBOR KEKEZ}
 
\address{Rudjer Bo\v{s}kovi\'{c} Institute,
         POB 1016, 10001 Zagreb, Croatia} 

\date{today}

\maketitle
\abstracts{
We present a recent treatment of the 
$\pi^0\gamma^\star\gamma$ transition form factor
in the coupled Schwinger-Dyson and Bethe-Salpeter approach to stress
the desirability of measuring it at CEBAF. 
}


The form factor $T_{\pi^0}(-Q^2,0)$ for the transition 
$\gamma^\star(k)\gamma(k^{\prime})\to\pi^0(p)$ (where
$k^{2} = - Q^2 \neq 0$ is the momentum-squared of the
spacelike off-shell photon $\gamma^\star$), is usually
thought to be almost completely understood on the basis 
of the perturbative QCD (pQCD) over the whole range of the 
momenta-squared covered by the new CLEO data \cite{gronberg98}
(1.5 GeV$^2 \lsim Q^2 \lsim 9$ GeV$^2$).
However, Radyushkin and Ruskov's \cite{Radyushkin+Rusk3}
QCD sum rule analysis indicates the presence of large 
nonperturbative contributions even for the largest of
these presently accessible momenta.
This motivated us to address 
in Ref. \cite{KeKl3} the transition form factor in the 
coupled Schwinger--Dyson (SD) and Bethe--Salpeter (BS)
approach in all spacelike-momentum regimes, from very 
large $Q^2$ all the way to $Q^2=0$. The $Q^2=0$ limit 
corresponds to the $\pi^0$ decay into two real photons 
($k^{2} = k^{\prime 2} = 0$)
explained by the Abelian axial {\it alias} 
Adler-Bell-Jackiw (ABJ) anomaly, which is usually difficult
to incorporate into a bound-state approach \cite{KeBiKl98}.
Fortunately, as shown already in Refs. \cite{bando94,Roberts} 
and used in the first analysis of $T_{\pi^0}(-Q^2,0)$ in a SD 
approach \cite{Frank+al}, such approaches which rely on the 
chirally well--behaved SD (and BS) equations for the light 
pseudoscalar mesons ($\pi, K, \eta, ...$) while respecting  
the Ward-Takahashi identities (WTI) of QED, 
reproduce (in the chiral and soft limit) 
the fundamental anomaly result 
\begin{equation}
 T_{\pi^0}(0,0) =\frac{1}{4\pi^2 f_\pi} \
\label{AnomAmpl}
\end{equation}
naturally and without any fine tuning.
This is an advantage with respect to pQCD approaches, which
have problems at low $Q^2$. ({\it E.g.}, see Ref. \cite{M+Rady97}.) 

In the present coupled SD-BS approach \cite{KeKl1,KlKe2,KeBiKl98}, 
the anomalous amplitude (\ref{AnomAmpl}) is also obtained 
model-independently and without imposing any requirements  
on the solutions for the $\pi^0$ wave function or 
the quark propagator solutions ({\it e.g.}, see 
\cite{bando94,Roberts}). This is exactly as it should be, since 
the anomaly -- and its result (\ref{AnomAmpl}) 
-- must not depend on the internal structure of the pion.
Subsequently, in Ref. \cite{KeKl3}, we showed that the SD-BS approach 
to modeling QCD provides a good description for {\it both} low and high 
values of $Q^2$. We thereby obtained the behavior 
similar to the Brodsky--Lepage interpolation formula 
$T_{\pi^0}(-Q^2,0)=(1/4\pi^2 f_\pi) \, / \, (1 + Q^2/8\pi^2 f_\pi^2)$
proposed \cite{BrodskyLepage} as a desirable behavior for 
$T_{\pi^0}(-Q^2,0)$
as it reduces to the ABJ anomaly amplitude (\ref{AnomAmpl}) at $Q^2=0$, 
while agreeing with the following type of leading behavior
for large $Q^2$:
\begin{equation}
 T_{\pi^0}(-Q^2,0) = {\cal J} \, \frac{f_\pi}{Q^2} \,
\qquad ({\cal J} = {constant} \,\, {\rm for \,\, large \,\,} Q^2),
\label{largeQ2}
\end{equation}
which is favored both experimentally \cite{gronberg98} 
and theoretically ({\it e.g.}, 
\cite{BrodskyLepage,manohar90,Radyushkin+Rusk3,M+Rady97}).

In the coupled SD-BS approach, the BS equation for the pion bound-state 
$q\bar q$ vertex $\Gamma_{\pi^0}(q,p)$ employs the dynamically dressed 
quark propagator $S(k) = [ A(k^2)\slashed{k} - B(k^2) ]^{-1}$, obtained 
by solving its SD equation. 
Then, in the case of light pseudoscalars, the $q\bar q$ bound 
states are simultaneously also the (pseudo-)Goldstone bosons of 
dynamical chiral symmetry breaking (D$\chi$SB).

Following Jain and Munczek \cite{jain91+munczek92,jain93b}, we adopt 
the ladder-type approximation 
employing bare quark--gluon--quark vertices but dressed propagators. 
For the gluon propagator we use an 
effective, ({\it partially}) modeled one in Landau-gauge 
\cite{jain91+munczek92,jain93b}, given by
$G(-l^2)( g^{\mu\nu} - {l^\mu l^\nu}/{l^2} )~.$
(This Ansatz is often called the 
``Abelian approximation" \cite{MarisRoberts97PRC56}.)
The effective propagator function $G$ is the
sum of the perturbative contribution $G_{UV}$ and the nonperturbative
contribution $G_{IR}$:
$G(Q^2) = G_{UV}(Q^2) + G_{IR}(Q^2)~,\;\;(Q^2 = -l^2)~.$
The perturbative part $G_{UV}$ is required to reproduce correctly
the ultraviolet (UV) asymptotic behavior that unambiguously
follows from QCD in its high--energy regime.
Therefore, $G_{UV}$ must be given by
$\alpha_s(Q^2)/Q^2$, where 
the running coupling constant $\alpha_s(Q^2)$ is
well-known from pQCD, so that $G_{UV}$ is {\it not} modeled.
As in Refs. \cite{KeKl1,KlKe2,KeBiKl98}, we follow 
Refs. \cite{jain91+munczek92,jain93b} and employ the 
two--loop asymptotic expression for $\alpha_s(Q^2)$.
For the modeled, IR part, we adopt from Ref.~\cite{jain93b}
$G_{IR}(Q^2)=(16\pi^2/3) \,a\,Q^2 e^{-\mu Q^2}$ 
and $a=(0.387\,\GeV)^{-4}$, $\mu=(0.510\,\GeV)^{-2}$.

Our calculational procedures are detailed in our 
Refs. \cite{KeKl1,KlKe2,KeBiKl98}. We reproduce the solutions of 
Ref. \cite{jain93b} for the dressed propagator $S(q)$, {\it i.e.},  
$A(q^2)$ and $B(q^2)$, as well as the solutions for the 
four functions comprising the pion bound-state vertex $\Gamma_{\pi^0}$. 
Actually, Ref. \cite{jain93b} employs the BS amplitude 
$\chi_{\pi^0}(q,p)\equiv S(q+{p}/{2})\Gamma_{\pi^0}(q,p)S(q-{p}/{2})$,
which is completely equivalent.

We assume that the $\pi^0\gamma^\star\gamma$ transition 
(and other pseudoscalar meson $\to$ two photon transitions)
proceeds through the pseudoscalar-vector-vector triangle 
graph (Fig.~1), and that we calculate the pertinent amplitude 
$T_{\pi^0}^{\mu\nu}(k,k^\prime) = \varepsilon^{\alpha\beta\mu\nu}
 k_\alpha k^\prime_\beta T_{\pi^0}(k^2,k^{\prime 2})~$
as in Refs. \cite{KeKl1,KlKe2,KeBiKl98},
using the framework advocated by, {\it e.g.}, 
Refs.~\cite{bando94,Roberts,Frank+al}
in the context of electromagnetic interactions of BS bound states,
and often called ({\it e.g.}, by Ref. \cite{Frank+al}) 
the generalized impulse approximation (GIA).
We therefore use the {\it dressed} quark propagator $S(q)$
and the pseudoscalar BS bound--state vertex $\Gamma_P(q,p)$,
as well as consistently dressed
{\it electromagnetic} vertex $\Gamma^\mu(q^\prime,q)$,
which satisfies the vector Ward--Takahashi identity (WTI)
$(q^\prime-q)_\mu \Gamma^\mu(q^\prime,q)=S^{-1}(q^\prime)-S^{-1}(q)~.$
\begin{figure}
\begin{center}
     \epsfxsize=9cm \epsfbox{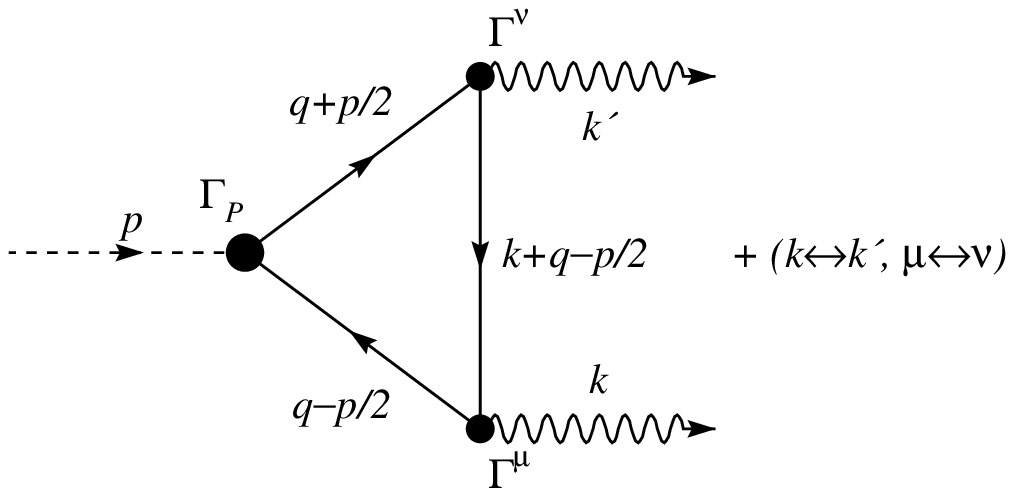}
      \caption{ The GIA diagram
                for $P^0 \to\gamma^{(\star)}\gamma^{(\star)}$ processes
                ($P^0 = \pi^0,\eta, \eta^\prime, \eta_c, \eta_b$).
                 }
      \label{triangle.graph}
\end{center}
\end{figure}                                                                  
Namely, assuming that photons
couple to quarks through the bare vertex $\gamma^\mu$
would be inconsistent with our dressed quark propagator $S(q)$, 
which contains the momentum-dependent functions $A(q^2)$ and $B(q^2)$.
The bare vertex $\gamma^\mu$ obviously violates the vector WTI, implying 
the nonconservation of the electromagnetic vector current and charge.
Solving the pertinent SD equation for the dressed quark--photon--quark 
($qq\gamma$) vertex $\Gamma^\mu$ is still an unsolved problem, and using 
the realistic Ans\"{a}tze for $\Gamma^\mu$ still remains the only practical 
way to satisfy the WTI.  The simplest particular solution of the vector WTI 
is the Ball--Chiu (BC) ~\cite{BC} vertex 
        \begin{eqnarray}
        \Gamma^\mu_{BC}(q^\prime,q) =
        A_{\bf +}(q^{\prime 2},q^2)
       \frac{\gamma^\mu}{\textstyle 2}
        + \frac{\textstyle (q^\prime+q)^\mu }
               {\textstyle (q^{\prime 2} - q^2) }
        \{A_{\bf -}(q^{\prime 2},q^2)
        \frac{\textstyle (\slashed{q}^\prime + \slashed{q}) }{\textstyle 2}
         - B_{\bf -}(q^{\prime 2},q^2) \}~,
        \label{BC-vertex}
        \end{eqnarray}
where
$H_{\bf \pm}(q^{\prime 2},q^2)\equiv [H(q^{\prime 2})\pm H(q^2)]$,
for $H = A$ or $B$.
It does {\it not} introduce any new parameters as it is completely
determined by the dressed quark propagator $S(q)$. 
In the SD-BS approach,
this minimal WTI-satisfying Ansatz (\ref{BC-vertex}) is 
still the most widely used $qq\gamma$ vertex
({\it e.g.}, Refs. \cite{Frank+al,Roberts,KeKl1,KlKe2}).
A general WTI-satisfying vertex can be written \cite{BC} as 
$\Gamma^\mu = \Gamma^\mu_{BC} + \Delta \Gamma^\mu$,
where the addition $\Delta \Gamma^\mu$ doesn't contribute
to the WTI, since it is transverse, 
$(q^\prime-q)_\mu \Delta \Gamma^\mu(q^\prime,q) = 0$.
That is, $\Delta \Gamma^\mu(q^\prime,q)$ lies
in the hyperplane spanned by the vectors ${\cal T}_i^\mu(q^\prime,q)$
($i = 1, ..., 8$) transverse to the photon momentum $k=q^\prime-q$.
Curtis and Pennington (CP) \cite{CP90} advocated an Ansatz for
$\Delta \Gamma^\mu(q^\prime,q)$ exclusively along the $i = 6$ basis 
vector: $\Delta \Gamma^\mu(q^\prime,q) = {\cal T}_6^\mu(q^\prime,q)
[A_{\bf -}(q^{\prime 2},q^2)/2 d(q^\prime,q)]$.
The coefficient of ${\cal T}_6^\mu(q^\prime,q)$ can be 
chosen to ensure renormalizability \cite{CP90}, {\it e.g.},
\begin{equation}
d_{\pm}(q^\prime,q) =
        \frac{1}{q^{\prime 2}+q^2}
        \left\{
                (q^{\prime 2} \pm q^2)^2
                +
                \left[ M^2({q^\prime}^2) + M^2(q^2) \right]^2
        \right\}~,
\label{defdfunct}
\end{equation}
where $M(q^2) \equiv B(q^2)/A(q^2)$ is the D$\chi$SB-generated
dynamical mass function, which in our case has the large-$q^2$ 
dependence \cite{KeBiKl98} in agreement with pQCD.

The choice $d = d_-$ corresponds to the CP vertex Ansatz
$\Gamma^\mu_{CP}$ \cite{CP90}. 
We use it in analytic calculations of $T_{\pi^0}(-Q^2,0)$, 
which are possible for $Q^2=0$ and $Q^2\to \infty$.
However, in the numerical calculations, which are 
necessary for finite values of $Q^2\neq 0$, we prefer the 
{\it modified} CP (mCP) vertex, $\Gamma^\mu_{mCP}$ \cite{KeKl3}, 
resulting from the choice $d = d_+$, since it is
easier to deal with numerically.
Unlike the BC one, the mCP vertex is consistent with 
renormalizability just like the CP one.
In the present context,
the important {\it qualitative} difference between the BC vertex, 
and the CP as well as the modified, mCP vertex, 
is that $\Gamma^\mu_{BC}(q^\prime,q)\to\gamma^\mu$
when {\it both} $q^{\prime 2},q^2\to \pm \infty$, whereas 
$\Gamma^\mu_{CP}(q^\prime,q) \to \gamma^\mu$ and 
$\Gamma^\mu_{mCP}(q^\prime,q) \to \gamma^\mu$ 
as soon as {\it one} of the squared momenta tends to infinity.

Assuming the isospin symmetry, 
$\chi \equiv \chi_{u\bar u} = \chi_{d\bar d}$ is the
BS amplitude for both $u\bar u$ and $d\bar d$ pseudoscalar
bound states, 
and $\chi_{\pi^0}(q,p) \equiv \chi(q,p) \, \lambda^3/\sqrt{2}$.
Then, for $\pi^0$, GIA yields (see Fig.~1) the tensor amplitude 
\begin{displaymath}
        T_{\pi^0}^{\mu\nu}(k,k^\prime)
        =
        -
        N_c \,
        \frac{1}{3\sqrt{2}}
        \int\frac{d^4q}{(2\pi)^4} \mbox{\rm tr} \{
        \Gamma^\mu(q-\frac{p}{2},k+q-\frac{p}{2})
        S(k+q-\frac{p}{2})
\end{displaymath}
\begin{equation}
          \qquad
        \times
        \Gamma^\nu(k+q-\frac{p}{2},q+\frac{p}{2})
        \chi(q,p) \}
        +
        (k\leftrightarrow k^\prime,\mu\leftrightarrow\nu),
\label{Tmunu(2)}
\end{equation}
leading to the analytical reproduction of the $\pi^0\to\gamma\gamma$ 
amplitude (\ref{AnomAmpl}) with the CP and mCP vertices, in the same 
way as with the BC vertices.

Let us stress that in the regime of asymptotically large $Q^2$,
we obtained \cite{KeKl3} analytical results for the transition
form factor in agreement with the behavior (\ref{largeQ2}).
Since the pion is light, $k\cdot k^\prime \approx Q^2/2$
for large negative $k^{2}=-Q^2$ and ${k^\prime}^2=0$.
Taking into account the behavior of the propagator functions 
$A(q^2)$ and $B(q^2)$ ({\it e.g.}, see \cite{KeBiKl98})
we can then in Eq. (\ref{Tmunu(2)}) approximate those quark propagators 
that depend on the photon momenta $k$ and $k^\prime$, by their  
asymptotic forms: $S(q-p/2+k) \approx S(k-p/2) \approx
-(2/Q^2)(\slashed{k}-\slashp/2)$
and analogously for the propagator where $k$ is replaced by $k^\prime$.
Although the relative loop momentum $q$ can be large in the course of 
integration, its neglecting is justified because 
the BS amplitude $\chi(q,p)$ decays quickly and thus strongly 
damps the integrand for the large $q'$s.
Ref. \cite{KeKl3} then showed that 
$T_{\pi^0}^{\mu\nu}(k,k^\prime) = -(2/N_c \, Q^2)\,
\varepsilon^{\mu\lambda\nu\sigma}(k-k^\prime)_\lambda p_\sigma \, f_\pi$,
finally leading to Eq. (\ref{largeQ2}) with the coefficient  
${\cal J}=4/3$. We thus found {\it model independently} 
that the asymptotic behavior in the present approach 
(with $qq\gamma$ vertices such as the bare $\gamma^\mu$, mCP and CP), 
agrees exactly with the leading term predicted by OPE \cite{manohar90}.

The asymptotic behavior $\propto 1/Q^2$ obtained in the present approach is 
especially satisfying when compared with the one known to result from the 
simple constituent quark model (with the constant light--quark mass parameter 
$m_u$), where $T_{\pi^0}(-Q^2,0) \propto (m_u^2/Q^2)\ln^2(Q^2/m_u^2)$
as $Q^2\to\infty$, which overshoots the data considerably
because of the additional $\ln^2(Q^2)$-dependence.

It is important to note that this asymptotic behavior, 
$T_{\pi^0}(-Q^2,0) = (4/3) f_\pi/Q^2$, was obtained for 
the $qq\gamma$ vertices which reduce to the bare one, $\gamma^\mu$,
as soon as {\it one} of the squared momenta tends to infinity,
such as the renormalizable CP and mCP vertices. Subsequently, 
Ref. \cite{RobertsDubr} showed that this asymptotic behavior 
must in fact hold for {\it any} renormalizable $qq\gamma$ vertex.
For the simplest dressed WTI-preserving vertex, the BC one, which
reduces to the bare one only when momenta are large in both of its 
fermion legs, and which is not consistent with multiplicative 
renormalizability \cite{CP90}, the asymptotic behavior
is $(4/3) {\widetilde f}_\pi/Q^2$, where the quantity 
${\widetilde f}_\pi$ is given by the same Mandelstam-formalism
expression as $f_\pi$ except that its integrand is modified by 
the factor $[1+A(q^2)]^2/4$ \cite{KeKl3}. For our model solutions 
\cite{KeBiKl98}, ${\widetilde f}_\pi \approx 1.334 f_\pi$, so that 
the usage of the BC vertices causes the increase of ${\cal J}$ in 
Eq. \ref{largeQ2} from ${\cal J}=4/3$ to ${\cal J}\approx 1.78$.

Of course, the model-independent asymptotic 
coefficient $4 f_\pi/3$ is the one having the more fundamental meaning, 
resulting from the renormalizable $qq\gamma$ vertices such as the CP or 
mCP ones, which have properties closer to the true vertex solution.  
Also indicative is the asymptotics found \cite{KeKl3} 
for such $qq\gamma$ vertices when both photons are 
off-shell, $k^2 = - Q^2 << 0$ and $k^{\prime 2} = - Q^{\prime 2} \leq 0$:
\begin{equation}
T_{\pi^0}(-Q^2, -Q^{\prime 2}) = \frac{4}{3} \,
                                \frac{f_\pi}{Q^2 + {Q^{\prime}}^2} \, .
                                \label{baregasgas}
\end{equation}
The usage of the BC vertex (\ref{BC-vertex}) would again modify
this result by the substitution $f_\pi \to {\widetilde f}_\pi$.
However, Eq.~(\ref{baregasgas}) agrees with the leading term of 
the OPE result of Novikov {\it et al.} \cite{novikov+al84} for 
the special case $Q^2=Q^{\prime 2}$. 
Also, the distribution-amplitude-dependence of the pQCD approach 
cancels out for that symmetric case, so that 
$T_{\pi^0}(-Q^2, -Q^{\prime 2})$ in this approach ({\it e.g.},
see \cite{Kess+Ong93}), in the limit $Q^2={Q^{\prime}}^2\to\infty$,
exactly agrees with both our Eq. (\ref{baregasgas}) and Ref.
\cite{novikov+al84}. For that symmetric case, we should thus
have even the precise agreement of the coefficients
irrespective of the description of the pion internal structure
encoded in the distribution amplitude. Obviously, this favors
the renormalizable $qq\gamma$ vertices, such as CP and mCP ones,
over the BC vertex. However, the BC vertex may anyway be the one 
which is more accurate not only for the presently accessible $Q^2$, 
but also for larger values before starting to fail.
\begin{center}
\mbox{\centering \epsfxsize=10cm \epsfbox{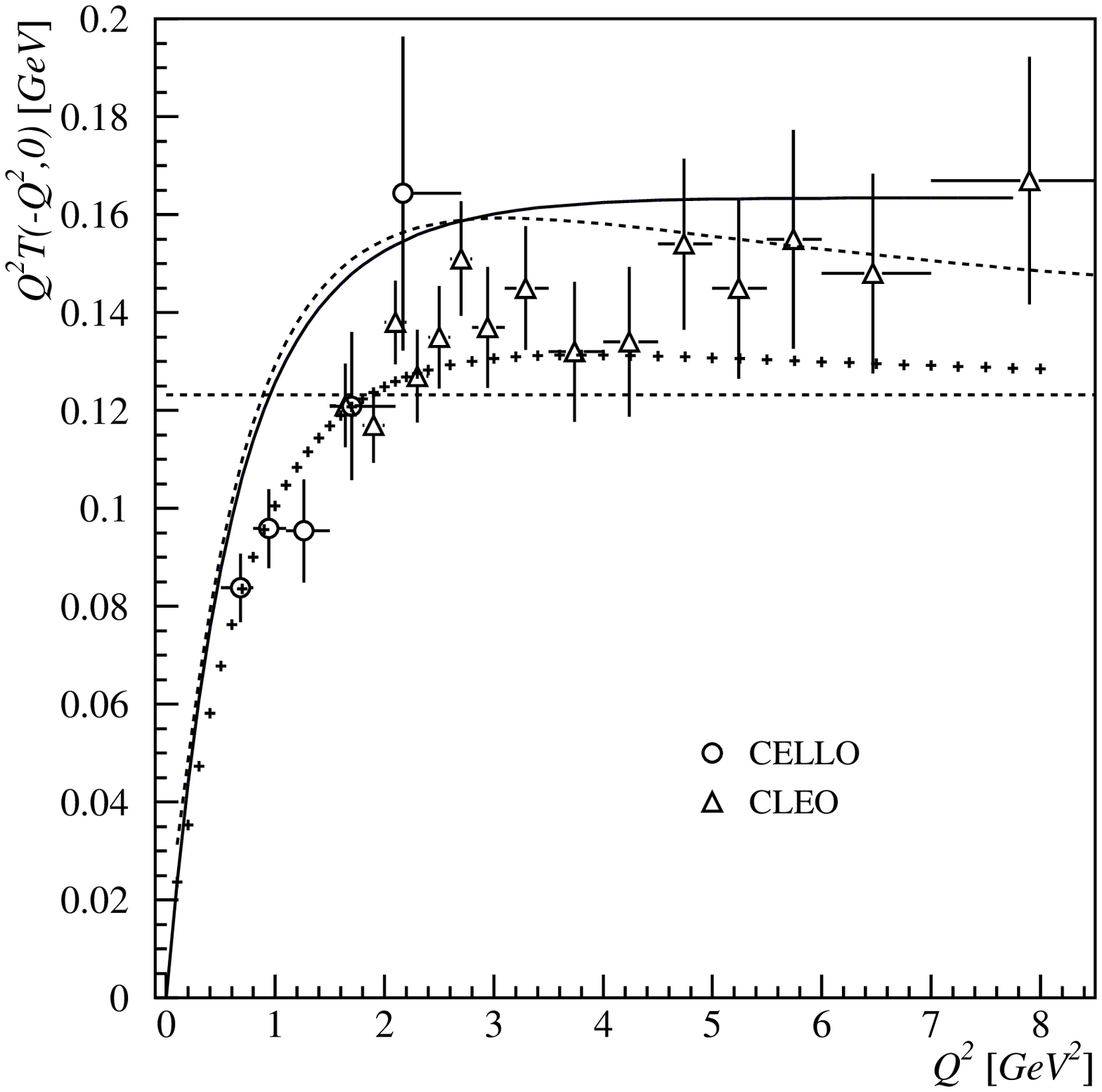}}
\end{center}
 
{\rm\small Figure 2: Our numerically obtained $Q^2 T_{\pi^0}(-Q^2,0)$
                     and the data.}

\vskip 4mm

For finite $Q^2$, the transition form factor 
is evaluated numerically. Fig.~2 compares so obtained 
$Q^2 T_{\pi^0}(-Q^2,0)$ with the CLEO \cite{gronberg98}
and CELLO \cite{behrend91} data. The solid curve is 
obtained with the BC vertices, and the dashed curve with
the mCP ones. The horizontal dashed line denotes $4 f_\pi/3$,
the asymptotic value for renormalizable $qq\gamma$ vertices 
such as the mCP one. It barely touches the experimental
error bars from below. Note that also Manohar \cite{manohar90} 
warned that his OPE approach (also yielding $4 f_\pi/3$)
indicates the possibility of large corrections to his leading 
term, but possibly also to the pQCD asymptotic coefficient
$2 f_\pi$, which is however exact in the 
{\it strict} $\ln(Q^2)\to\infty$ limit \cite{BrodskyLepage}. 
We get it lower by the factor 2/3 because our present approach
does not incorporate the effects of the pQCD evolution. 
However, as Ref. \cite{Radyushkin+Rusk3} seems to indicate
that effects other than the pQCD evolution (including the 
presently interesting dynamical dressing of quarks) may still 
play important (and maybe even dominant) role even at 
$Q^2$-values larger than the presently accessible ones, 
one should consider seriously also nonperturbative approaches 
such as ours. 
On the other hand, high-precision measurements of 
$T_{\pi^0}(-Q^2,0)$ can test (and give a hint on how
to improve) the SD and BS model solutions which have so far been
successful in fitting the low-energy hadron properties such as
the meson spectrum. For example, our model choice \cite{jain93b}
somewhat overshoots (for both BC and mCP vertices) the present
$T_{\pi^0}(-Q^2,0)$ data \cite{gronberg98,behrend91}
in the region $Q^2 \lsim 4$ GeV$^2$.
 
Once one has a solution for $\chi(q,p)$ that leads to the correct 
value of $f_\pi$, the transition form factor is most sensitive on 
$A(q^2)$, or, more precisely, on its values at small and intermediate 
momenta $-q^2$, where $A(q^2)$ is still appreciably different from 1
\cite{KeKl3,KlKe4}. To illustrate what happens when the
$A(q^2)$-profile is decreased, let us enforce by hand the extreme,
artificial case $A(q^2)\equiv 1$. (To avoid confusion, we stress it
is for illustrative purposes only, as we cannot have such a SD-solution
in the adopted approach of Refs.~\cite{jain91+munczek92,jain93b}.)
This leads to the curve traced on Fig.~2 by small crosses,
pertaining to the usage of both BC and mCP vertices (as well
as the CP ones), since
$A(q^2)\equiv 1$ makes $ \Gamma^\mu_{mCP} \to \Gamma^\mu_{BC}$.
This curve reveals how the heights of the curves depicting
$Q^2 T_{\pi^0}(-Q^2,0)$ depend on how much the $A(q^2)$-profile
exceeds 1. Obviously, for both the solid curve and the dashed one,
the agreement with experiment would be improved by lowering them
somewhat (at least in the momentum region $Q^2\lsim 4$ GeV),
which could be achieved by modifying the model \cite{jain93b} and/or its
parameters so that such a new solution for $A(q^2)$ is somewhat lowered
towards its asymptotic value $A(q^2\to\infty) \to 1$. (Of course, in
order to be significant, this must not be a specialized re-fitting aimed
{\it only} at $A(q^2)$. Lowering of $A(q^2)$ should be a result of a 
broad fit to many meson properties, comparable to the original 
fit \cite{jain93b}. This, however, is beyond the scope of this paper.)                           

By the same token, high precision measurements of $T_{\pi^0}(-Q^2,0)$
can be especially helpful in obtaining information on which solutions 
for $A(q^2)$ are empirically acceptable and which are not. Of course, 
measurements of $T_{\pi^0}(-Q^2,0)$ give information on the integrated 
strength of $A(q^2)$ rather than on $A(q^2)$ itself. However, since it 
is known that the form of that function must be a rather smooth 
transition ({\it e.g.}, see \cite{KeBiKl98})
from $A(q^2)>1$ for $q^2$ near 0, to $A(q^2)\to 1$ in the $q^2$-domain
where QCD is perturbative, such measurements \cite{gagasCEBAF} would
give a useful hint even about $A(q^2)$ itself -- namely about what
solutions for $A(q^2)$ one may have in sensible descriptions of
dynamically dressed quarks and their bound states.
Therefore, the intermediate-momentum ($Q^2 \lsim 4$ GeV$^2$)
measurements of $T_{\pi^0}(-Q^2,0)$ at Jefferson Lab,
such as those proposed in Ref. \cite{gagasCEBAF}, would be very 
desirable. Hopefully, the study of virtual Compton scattering at
CEBAF can produce the data on $T_{\pi^0}(-Q^2,0)$ as envisioned in  
Ref. \cite{gagasCEBAF}.


\section*{Acknowledgments}
\noindent D. Klabu\v{c}ar thanks the organizers C. Carlson and
A. V. Radyuskin for the wonderful workshop. The authors
acknowledge the support of the Croatian Ministry of Science 
and Technology. 


\section*{References} 
\begin{thebibliography}{10}

\bibitem{gronberg98}
J. Gronberg {\it et al.} (CLEO collaboration),
Phys. Rev. D {\bf 57}, 33 (1998).

\bibitem{Radyushkin+Rusk3}
A. V. Radyushkin and R. T. Ruskov,
Nucl. Phys. B {\bf 481}, 625 (1996).
 
\bibitem{KeKl3}
D. Kekez and D. Klabu\v car,
Phys. Lett. {\bf 457B}, 359 (1999).
                                                                               
\bibitem{KeBiKl98}
D. Kekez 
{\it et al.}, Int. J. Mod. Phys. A {\bf 14}, 161 (1999).                                    

\bibitem{bando94}
M. Bando, M. Harada, and T. Kugo, Prog. Theor. Phys. {\bf 91}, 927 (1994).
 
\bibitem{Roberts}
C.~D. Roberts, Nucl. Phys. A {\bf 605}, 475 (1996).

\bibitem{Frank+al}
M.R. Frank 
{\it et al.}, Phys. Lett. B {\bf 359}, 17 (1995).
                                                                                \bibitem{M+Rady97}
I. V. Musatov and A. V. Radyushkin,
Phys. Rev. D {\bf 56}, 2713 (1997).
                                                                                
\bibitem{KeKl1}
D. Kekez and D. Klabu\v car, Phys. Lett. B {\bf 387}, 14 (1996).
 
\bibitem{KlKe2}
D. Klabu\v car and D. Kekez, Phys. Rev. D {\bf 58}, 096003 (1998).
                                                                                
\bibitem{BrodskyLepage}
G. P. Lepage and S. J. Brodsky, Phys. Rev. D {\bf 22}, 2157 (1980).
                      
\bibitem{manohar90}
A. Manohar, Phys. Lett. {\bf B244}, 101 (1990).
                                                                          

\bibitem{jain91+munczek92}
P. Jain and H.~J. Munczek, 
Phys. Rev. D {\bf 44}, 1873 (1991); Phys. Rev. D {\bf 46}, 438 (1992).

\bibitem{jain93b}
P. Jain and H.~J. Munczek, Phys. Rev. D {\bf 48}, 5403 (1993).


\bibitem{MarisRoberts97PRC56}
P. Maris and C.~D. Roberts, Phys. Rev. C {\bf 56}, 3369 (1997).


\bibitem{BC}
J. S. Ball and T.-W. Chiu, Phys. Rev. D {\bf 22}, 2542 (1980).

\bibitem{CP90}
D. C. Curtis and M. R. Pennington, Phys. Rev. D {\bf 42}, 4165 (1990).

\bibitem{RobertsDubr}
 C.~D.~Roberts, Fizika B (Zagreb) {\bf 8}, 285 (1999).                          

\bibitem{novikov+al84}
V. A. Novikov 
{\it et al.}, Nucl. Phys. B {\bf 237}, 525 (1984).

\bibitem{Kess+Ong93}
P. Kessler and S. Ong,
Phys. Rev. D {\bf 48}, R2974 (1993).

\bibitem{behrend91}
H.-J. Behrend {\it et al.} (CELLO Collaboration),
Z. Phys. C {\bf 49}, 401 (1991).
                                                                                
\bibitem{KlKe4}
D. Klabu\v car and D. Kekez, Fizika B (Zagreb) {\bf 8}, 303 (1999).
                                                                                
\bibitem{gagasCEBAF}
CEBAF Letter of Intent \# LOI--94/005. Co--Spokesmen: A. Afanasev,
J. Gomez, and S. Nanda; A. Afanasev, {\tt hep-ph/9608305}.
             
\end{thebibliography}
\end{document}